# Observability and Computability in Physical Systems

Subhash Kak

*Abstract:* This paper considers the relevance of the concepts of observability and computability in physical theory. Observability is related to verifiability which is essential for effective computing and as physical systems are computational systems it is important even where explicit computation is not the goal. Specifically, we examine two problems: observability and computability for quantum computing, and remote measurement of time and frequency.

**Introduction**
A system is said to be observable if a finite set of measurements on it completely determines the state of the system [1]. Less formally, the internal states of an observable system can be inferred from its behavior. Computability, on the other hand, is the ability to solve a problem effectively. In the absence of observability, a problem may not be computable if the unknown internal states can have influence on the computation. In addition, computability may not be achievable due to the logical nature of the problem [2].

Although the concepts of observability and computability are fundamental to system theory they are not explicitly considered in physics for the goals of the physicist are often framed in a narrower sense that involves observation of certain variables and not the knowledge of the complete state of the system. In certain cases the problem is merely to determine whether a specific state exists, as in the case of neutrino oscillations taken to be a consequence of the superposition of three neutrino states of different mass.

The classical-quantum dichotomy itself defines two distinct domains in which the amount of knowledge that can be gained about the system under study is different. The question if quantum description of physical reality is complete is the subject of an old debate [3],[4] and this question has bearing on the notion of observability. If the physical description of matter is complete there are no hidden states to be discovered. The generally accepted resolution of this debate is based on experimental results related to the Bell inequality [5].

The condition of observability must be modified for quantum systems since interacting with a quantum state causes it to collapse and the interaction, furthermore, is circumscribed by the uncertainty principle. Rather than make measurements on a single state, identical copies of it need be examined to determine it [6]. Single quantum states can, in principle, be prepared, but any such state cannot be verified with complete certainty. If quantum theory must deal with questions of completeness owing to its non-realistic interpretations that lead to paradox [7], there is search for more fundamental models in other areas of physics that is motivated either by unification of forces or resolution of experimental discrepancies [8],[9]. Some models assume that matter in its ordinary states is not fully observable and hidden variables will be revealed as energies are raised many-fold.

If incompleteness is inevitable in the consideration of properties of elements of a set and for a formal system, as given by Gödel's theorem, paradox is inevitable in the consideration of observers due to self-referral. These difficulties arise from the conception of being who enters the discourse through the agency of the observer. For the same reason, problems of observability and computability cannot be avoided in considerations of the universe. Given that, by definition, the universe is all there is, its descriptions must suffer from problems of self-referral with attendant paradox and they will admit different interpretations [10],[11].

This paper examines the problems of observability and computability in the contexts of quantum computing and remote measurement of time and frequency. As observability cannot be guaranteed in quantum computing systems, claims regarding their computational use must be questioned. It is implicitly assumed that questions of observability and computability do not arise for a simple classical system if the measurements have sufficient precision but we show that this assumption is invalid in certain situations where the system is remotely located. Time and frequency associated with systems also do not always satisfy the observability criterion. This is surprising since noninertial frames are generally associated with non-uniform motion which, one would assume, can be easily established.

**States and their Verification**
The term *information* has a variety of meanings. In the mathematical theory of communications, information is a measure of the surprise associated with the received signal. It is implied that the receiver has knowledge of the statistics of the messages produced, or to be produced, by the sender. A more likely signal carries less information as it comes with less surprise. This assumes that the system is observable.

In a classical system, the sender and the receiver share a set of messages from a specific alphabet and the statistics of the communications make it possible to determine the probability of each letter of the alphabet. The information measure of the message $x$ associated with probability $p_x$ is $-\log p_x$ and the entropy, or average information of the system, is given by $H(X) = -\sum_x p_x \log p_x$. The amount of information associated with an object could be taken to mean the amount necessary to completely describe it. Since this information will vary depending on the interactions the object has with other objects and fields, and thus be variable, it may be measured for the situation where the object is isolated.

In the quantum context, the question of state is more complex than in the classical case. The quantum state cannot be completely isolated if it is entangled. The state cannot be known from a single object but rather is a representation of our knowledge of the system. Such knowledge can only be assembled based on several copies of the system. One cannot consider the actual computation in abstract mathematical terms for it is ultimately a physical process [12],[13],[14]. The amount of information that can be obtained from many copies of the unknown quantum state is different from the von Neumann measure [6] which is used to characterize this information.

An effective computation requires that the initial state of the system be completely known, which is not always possible [15]. In reality initial states as well transformations will have errors. Furthermore, if the computation involves entangled states, there is no way any specific pair of states can be verifiably known to be so. Even when many copies of the quantum state are available, it may not be known whether this quantum state is entangled with some other states [16].

As example, consider teleportation that requires sharing of an entangled pair of states by two individuals who are separated in distance. But whether a given pair of states is entangled cannot be verified by any specific tests. Therefore, one can never prove that an unknown quantum state has been successfully teleported.

The states in a classical computation may be verified by carrying out another computation on identical equipment, which is possible since the computation is discrete (that is associated with threshold phenomenon that suppresses small errors). In quantum computing, the computation is analog and therefore copies of the computation will not be identical. In any non-trivial computation, the small differences (in each step) will accumulate so as to make the overall computation unverifiable. Furthermore, interaction with the environment would lead to decoherence.

**Inertial and Noninertial Frames**

In Newtonian mechanics, an inertial frame is a reference-frame relative to which the motion of a body not subject to forces is always rectilinear and uniform. Any other frame of reference moving uniformly relative to an inertial frame is also an inertial frame. In electrodynamics, an inertial frame is one in which light travels with the same speed in all directions. If one holds on to absolute simultaneity, then clearly another frame that moves uniformly relative to such a frame will not satisfy the definition.

By definition, inertial frames are equivalent and laws have the same form in them. The presence of fields requires changes in the form of the laws and there are further complications due to violation with respect to parity and time-reversal [17]. Yet, it is convenient to begin with the assumption that the laws are the same everywhere in the universe and that is why in cosmological considerations objects are taken to be inertial.

Noninertial frames are not associated with the same clock as inertial frames as evidenced by the traveling twin in the twin paradox who ages less than the stay-behind twin. With respect to the inertial frame, the clock on the noninertial frame will be slower.

Since the frames can, in principle, calibrate their clocks with respect to some other independent process, like the rotation period of a star, it follows that this could be used by a frame to discover its noninertial nature even if its speed at the time of the calibration were to be uniform.

Different clock rates also arise due to gravity. Clocks at higher gravitational potential run faster than those at lower gravitational potential [18]. There is further slowing of the clocks due to the cosmological expansion of the universe. But these non-kinematic factors do not concern us here. If one were to define inertiality with respect to the large scale structure of the universe [19],[20], the perspective of the noninertial frames, with their varying clock rates, becomes important.

For a universe where clocks run at different rates that are unknown to other observers, the unknown clock rates must be estimated to find distances that separate them. If the observers are too far from each other to be able to use radar ranging to estimate these distances, they can only communicate in one direction and they must form their estimates only by means of signals they receive from each other.

On the other hand, given relative Doppler shift values between frames, it is not always possible to determine if that shift has a component of noninertiality associated with it. Noninertial frames produce a frequency shift that is different from the Doppler shift of inertial frames. Due to the slowness of their clocks, noninertial frames show a larger redshift and if some of the light sources in the cosmos are noninertial then their redshift should be corrected. These corrections could be relevant for observed anomalies in redshifts [21].

**Slow Clocks and Doppler Shift**

According to Lorentz equations, moving clocks run slow by the factor $\sqrt{1 - v^2/c^2}$, with $v/c=\beta$, where $v$ is the relative velocity between the clock and a reference inertial frame and $c$ is the speed of light. For inertial frames this slowness is only apparent, and necessary for reconciling different viewpoints.

Assume that the observation is being made on inertial frame A with respect to which frame B is moving away with velocity *v*. The time on the frame of the clock, $t_B$, on B is related to the time, $t_A$, on the reference frame A by the relation:

$$t_B = t_A \sqrt{1 - v^2/c^2}$$

For light signals from frame B that are associated with frequency of $f_s$, the time duration between two crests of the light waves emitted is *1/$f_s$*. But during this time, the frame itself moves a distance of *v/$f_s$* and the time corresponding to that is $\dfrac{v}{f_s(c-v)}$ to give us a total time in the calculation with respect to frame B of:

$$\frac{1}{f_s} + \frac{v}{f_s(c-v)}$$

This time duration maps into $\left(\dfrac{1}{f_s} + \dfrac{v}{f_s(c-v)}\right)\sqrt{1 - v^2/c^2}$ with reference to the time of frame A and, therefore, this should equal the received time between crests equal to *1/$f_r$*. Simplifying,

$$\frac{1}{f_r} = \frac{1}{f_s}\sqrt{\frac{c+v}{c-v}}$$

The frequency of the received waves, $f_r$, is related to the frequency of the transmitted waves, $f_s$, by:

$$f_r = f_s \sqrt{\frac{c-v}{c+v}} = f_s \sqrt{\frac{1-\beta}{1+\beta}}$$

One can likewise compute the relationship between transmitted and received frequency and transverse motion between the frames where $\theta$ is the angle between the line joining the two frames and the direction of motion (Figure 1):

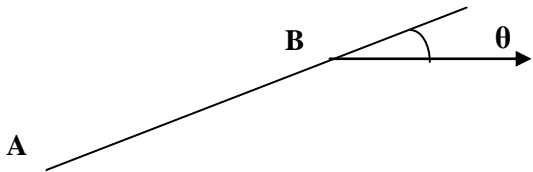

Figure 1

$$f_r = f_s \frac{1 - \frac{v}{c}\cos\theta}{\sqrt{1 - v^2/c^2}}$$

For noninertial frames the slowing of clocks is real. If two frames have different clock rates κ and λ, then κ wavelengths of one frame correspond to λ wavelengths of another frame.

Slow clocks mean that the natural frequencies of the processes have been scaled down on the frame compared to an inertial frame. If the scaling factor is α, the transmitted frequencies for natural processes would be scaled down by the factor α. The characteristic frequency of $f_s$ will be mapped into the slow-clock frame characteristic frequency of $\alpha f_s$.

**Apparent Velocity and Doppler Correction**

The observer cannot distinguish between the effect of a slow clock and velocity in the observations of the electromagnetic energy coming from a remote frame.

Assume B's characteristic frequency $f_B$ equals $\alpha f_A$. If A did not know that B had a different clock, it will assume that B was moving away from it with an apparent velocity $v_o$ and its characteristic frequency was no different from its own. Given these facts,

$$\alpha f_A = f_A \sqrt{\frac{c - v_o}{c + v_o}}$$

Simplifying, we get:

$$v_o = \frac{c(1 - \alpha^2)}{(1 + \alpha^2)} \qquad (1)$$

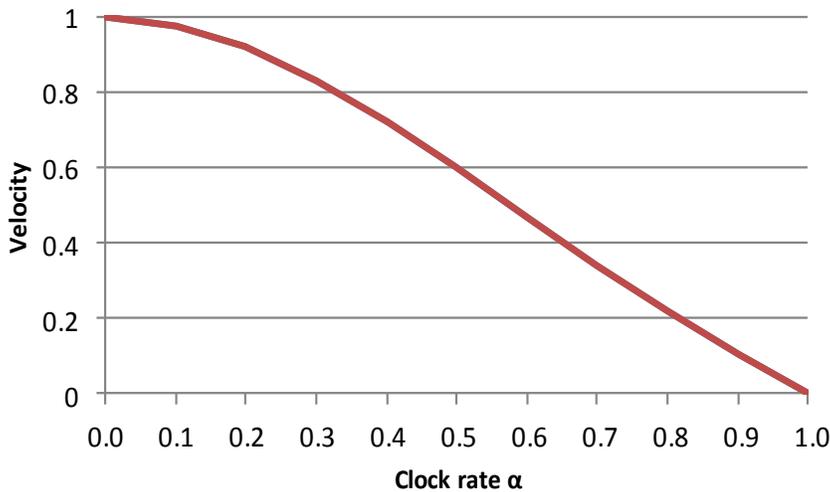

**Figure 1.** Relationship between clock rate α and apparent velocity $v_o$

Now we consider a more general situation where frame B (with its slow clock rate characterized by α) is moving away from A with a velocity of *u*. The characteristic frequency $f_r$ as received by A is given by:

$$f_r = \alpha f_A \sqrt{\frac{c-u}{c+u}}$$

Observers on frame A would deduce the velocity by the Doppler and begin by assigning frame B the velocity *v* according to:

$$f_r = \alpha f_A \sqrt{\frac{c-u}{c+u}} = f_A \sqrt{\frac{c-v}{c+v}} \qquad (2)$$

It follows that

$$\alpha^2 (c-u)(c+v) = (c-v)(c+u) \qquad (3)$$

If the value of *u* is known, then *v* depends on α according to the following relation:

$$v = \frac{c(c+u) - c(c-u)\alpha^2}{(c+u) + (c-u)\alpha^2} \qquad (4)$$

When $u = c(\alpha^2 - 1)/(\alpha^2 + 1)$, *v=0*. This represents the case where actual frame velocity is estimated to be zero by the observer. On the other hand, when *u=c, v=c,* or in the limiting case there is no error.

**Noninertial Frame with Clock Determined by Velocity Alone**
The ideas presented in this paper can be put to experimental test. For a clock that is on a noninertial frame moving at high velocity with respect to Earth (which is taken to be inertial), the estimated Doppler values would be different from the measured ones due to its inherent slow clock.

$$f_r = \alpha f_A \sqrt{\frac{c-v}{c+v}} = f_A \sqrt{\frac{c-v}{c+v}} \sqrt{1 - v^2/c^2}$$

In other words, the signals from this frame received on Earth will be associated with the frequency:

$$f_r = f_A (1 - v/c) \qquad (5)$$

This is the familiar formula for classical Doppler Effect. This means that noninertial frames are governed by classical Doppler formula even when the velocities under consideration are high.

For motion in the radial direction, the redshift formula for inertial frames ($z_I$) is:

$$z_I = \sqrt{\frac{c+v}{c-v}} - 1 \tag{6}$$

For the general noninertial case, this will have to be replaced by:

$$z_{NI} = \alpha^{-1}\sqrt{\frac{c+v}{c-v}} - 1 \tag{7}$$

For the specific case of formula (5), the corresponding expression is:

$$z_{NI} = \frac{v}{c-v} \tag{8}$$

We stress that the experiments in support of relativistic Doppler Effect (for example [22]) do not cover the noninertial cases described here.

For a comparison of the redshift values obtained using formulas (6) and (8), see Table 1. The larger values of redshift for noninertial frames is due to their intrinsic slow clock and for a proper comparison of their velocity, the corresponding figure in the same row for inertial frame redshift should be considered. In other words, if the observed redshift is, for example, 9.0, it should be replaced by the value of 3.36.

**Table 1.** Comparison of redshift for inertial and noninertial frames

| Velocity $v$ (fraction of $c$) | Inertial frame redshift $z_I$ | Noninertial frame redshift $z_{NI}$ |
|---|---|---|
| 0.1 | 0.11 | 0.12 |
| 0.2 | 0.22 | 0.25 |
| 0.3 | 0.36 | 0.43 |
| 0.4 | 0.53 | 0.66 |
| 0.5 | 0.73 | 1.00 |
| 0.6 | 1.00 | 1.50 |
| 0.7 | 1.38 | 2.66 |
| 0.8 | 2.00 | 4.00 |
| 0.9 | 3.36 | 9.00 |

We see that for higher values, the ratio of the redshifts for noninertial and inertial frames becomes progressively larger.

**Discussion**

The paper considered the problem of observability and computability in physical systems. It was argued that for quantum computing verifiability of the computation is not always possible even if one were to ignore questions of decoherence and noise. In kinematics it was concluded that the knowledge that can be inferred from observations may not, in general, help one choose between inertiality and noninertiality.

Noninertial sources have slow clocks and thus have intrinsic redshift even in the absence of motion. We obtained the surprising result that certain noninertial frames are governed by classical Doppler formula. If quasars and other sources with high redshift values [23] can be considered to be noninertial, the analysis in this paper has implications for estimates of their distance and speed.